\documentclass{elsart1p}
\usepackage{graphicx}

\begin{document}

\begin{frontmatter}

\title{Nonlinear Dynamic Phenomena in Macroscopic Tunneling}
\author{G. Dekel$^1$, O.V. Farberovich$^2$, A. Soffer$^3$ , V.
Fleurov$^1$}

\address{$^1$Raymond and Beverly Sackler Faculty of Exact Sciences,
School of Physics and Astronomy,\\ Tel-Aviv University, Tel-Aviv
69978 Israel.\\ $^2$ Department of Physics, Ben-Gurion University,
Beer-Sheva 84105, Israel.\\ $^3$Department of Mathematics, Rutgers
University, New Brunswick, NJ 08903,USA}

\begin{abstract}
Numerical simulations of the NLSE (or GPE) are presented
demonstrating emission of short pulses of the matter (light) density
formed in the course of tunneling in wave-guided light and/or
trapped BEC. The phenomenon is observed under various conditions,
for nonlinearities of different signs, zero nonlinearity included.
We study, both numerically and analytically, pulsations of matter
(light) remaining within the trap and use the results in order to
induce emission of sequential pulses by properly narrowing the trap.
This allows us to propose a mechanism for a realization of Atom
Pulse Laser.
\end{abstract}

\end{frontmatter}
\section{Introduction}

The research on dynamics of systems described by Non Linear
Schr\"odinger equation (NLSE), such as weakly interacting Bose -
Einstein condensates (BEC) treated in the mean field approximation
where it is known as Gross-Pitaevskii Equation (GPE), and light
propagating in nonlinear refractive media, has generated many
interesting results, most of which have to do with 'solitonic'
behaviors of these systems. The realization of BEC in 1995 has made
it possible to trace bright and dark matter wave solitons
experimentally \cite{becdark1,becdark2,becsol1,becsol2}, and
enhanced further vast theoretical work on the subject. The
similarity between the two physical objects, optical systems and
matter waves, both described by the similar equations, has made it
possible to study both systems in parallel and shed light on one
system by performing experiments on the other. Furthermore, the Atom
Optics field has emerged, and very interesting applications of the
optics principle implemented for matter waves were suggested and
realized\cite{atop1,atop2,atop3,atop4,atop5}. Interesting mechanisms
for realization of pulsed Atom laser were also suggested.
Refs.\cite{pulsedlaser1} and then \cite{pulsedlaser2} propose to use
spatial variation of the scattering length so that the trapped
condensate is accelerated towards a new minimum thus having a large
enough energy to escape from the trap as a single soliton. Partial
outcoupling is then achieved through an additional local maximum
which causes splitting in the attractive interatomic interaction
region, thus giving rise to a multiple emission. The authors also
note that the velocities of the ejected solitons are constant and
almost identical. Another mechanism was proposed in ref.
\cite{pulsedlaser3} where an elongated, quasi 1d repulsive
condensate is subject to simultaneous variations of the harmonic
trap, from attractive to repulsive, and of the interatomic
interaction, from repulsive to attractive, resulting in self
coherent solitonic pulses. A special attention is given there to the
stability criteria, mainly for avoidable collapse in the realistic
multidimensional case. In refs. \cite{pulsedlaser4,pulsedlaser5} a
periodic soliton array generator was introduced in the model that
contained two tunnel-coupled parallel cigar-shaped traps, in which
one trap is a BEC reservoir and the other trap is a lasing cavity.

Theoretical studies on dynamics of such systems have been carried
out both numerically, perhaps the most common is the Split Step
Fourier (SSF) method, and analytically, where some thorough and
fruitful investigation were carried out
\cite{fetter,stringani,anal}. Tunneling in one and two dimensional
systems in the WKB approximation (with and without vortices) was
addressed in the recent papers \cite{chm05,mcmb04}. Behavior of BEC
under the action of a time dependent field and tunneling through
chaotic areas was analyzed in ref. \cite{om05}. The hydrodynamic
formulation of the Schr\"odinger equation was originally proposed in
ref. \cite{made} and has been in use since both in quantum mechanics
and optics (see, e.g.
\cite{marbu,silber,ablo,Holland,stringani2,LF01}). An analytical
study of macroscopic tunneling dynamics within the hydrodynamic
representation of NLSE has been carried in Ref. \cite{fs05,ours}.
This study employs the adiabatic approximation making use of the two
characteristic time scales typical of the macroscopic tunneling
processes\cite{tunntime1,tunntime2,tunntime3,tunntime4}, and results
in an almost complete solution for the dynamics of tunneling of
weakly interacting BEC. The most important feature of this solution
is the observation of a tendency to formation of a dispersive shock
wave in the outskirts of the potential trap, which propagates later
on as a blip or pulse in the matter density or light intensity and
which under certain conditions may form a bright soliton. It is
important to emphasize that such pulses appear irrespective of the
nonlinearity in the NLSE, and the show up, in particular, in the
case of zero interaction (or zero Kerr nonlinearity in Optics).

Approaching BEC as a quantum fluid has recently given shock wave
dynamic phenomena its rightfully deserved attention. Dispersive
shock waves and related behaviors in BEC dynamics have been
predicted theoretically and observed experimentally. In classical
compressible gas and fluid dynamics, shock waves are known to be
traveling fronts of steep gradients. However, since GPE admits no
dissipation effects, the concept of dispersive shock waves has been
introduced.\cite{shock1,shock2,shock3,shock4,shock5} These are waves
that are generated and maintained by dispersion instead of
dissipation and are believed to consist of a number of spatial dark
solitons and exhibit large amplitude oscillations.

In this work we further investigate the dynamics of 1-d macroscopic
tunneling by simulating temporary and spatial evolution of tunneling
matter wave and light described within the frame work of NLSE/GPE.
We carry out detailed simulations which demonstrate an agreement
with formation and propagation of a blip \cite{ours} regardless of
the type of nonlinearity. However, the latter causes differences in
the further behavior of the outgoing pulse in the three cases,
negative, zero and positive nonlinearity, which correspond to
focusing, defocusing and 'nonfocusing' media in Optics, and
respectively attractive, repulsive and zero interactions in BEC.

We also demonstrate how different initial states rapidly transform
into the state optimal for the blip formation used in Ref.
\cite{ours}. This indicates that the process of blip formation is
not really sensitive to the initial conditions. We then concentrate
on the matter (light) which remains trapped after the emergence of
the blip, which is seen, both numerically and theoretically, to
pulsate with the frequency mainly determined by the trap.

In the last part of this work we merge the two studied effects, i.e.
single blip emission and periodic pulsations, to search for
conditions under which blips may be repeatedly and controllably
emitted. Such 'blip train' might pave a road for realization of Atom
Pulsed Laser.(\cite{pulsedlaser1, pulsedlaser2,
pulsedlaser3,pulsedlaser4,pulsedlaser5}) We show how each parameter
of such train is controllable.

\section{Simulation}\label{simulation}

\begin{figure}[tbp!]
\includegraphics[width=10cm]{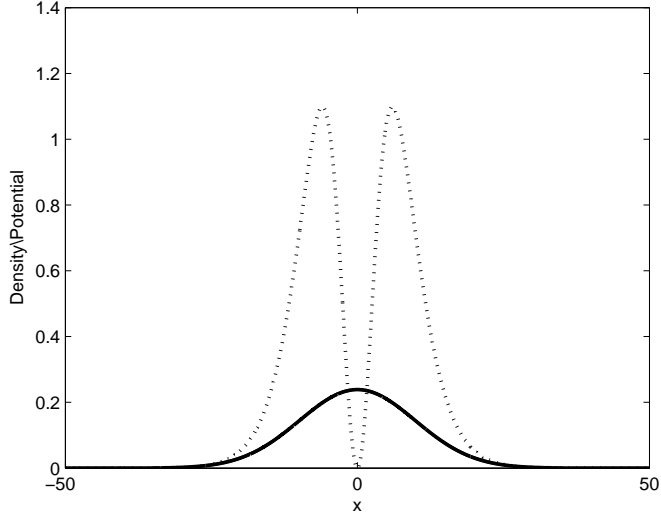}
%
%
\caption{The tunnel trap (dotted line) and initial packet (solid
line) for the parameters $C=0.1$, $a=5$, and $b=1$, and
$w=7$.}\label{struct}
\end{figure}

Our simulation program is based on the Interaction Pictures
Runge-Kutta-4 method \cite{Ballagh} for solving the dimensionless
GPE
\begin{equation}\label{GPE}
i\frac{\partial\Psi(x,t)}{\partial t}=\left[-\frac{1}{2}\nabla^2 +
U_{ext}(x) + NU_{int} \mid \Psi(x,t) \mid^2\right]\Psi(x,t)
\end{equation}
where $\Psi(x,t)$ is the normalized to one wave function of the
condensate with $N$ particles in it. The external potential
\begin{equation}\label{poti}
U_{ext} = C \frac{1}{\cosh^2(\frac{x}{a})}
\left(\frac{x}{b}\right)^2.
\end{equation}
is a symmetric tunnel trap (see Fig. \ref{struct}) parameterized by
the widths $a$ and $b$. $C$ is a constant which detemines the height
$U_0 = \mbox{max}[ U_{ext}(x)] \simeq1$ fo the trap. $U_{int}$ is
the effective interaction potential which may be positive
(repulsive) or negative (attractive), or even zero.
Our calculations use the values $\frac{NU_{int}}{U_0}= \pm 0.15,\
0$. Then the initial normalized Gaussian packet,
$$
\Psi_0(x) = \sqrt{\frac{1}{\sqrt{2\pi}w}}
\exp\left({-\frac{x^2}{4w^2}}\right),
$$
whose peak is situated at the center of the well at the time $t=0$,
evolves dynamically according to the GPE equation (\ref{GPE}). The
parameters of the trap are $C = 0.1$ (so that $U_0$ is close to
one), $a=5$, $b=1$, and the width of the initial Gaussian input is
$w=7$. Fig. \ref{struct} presents a sketch showing the trap and the
initial packet.

\subsection{Application to Optics}

In parallel, another simulation program, the Free Beam Propagation
Method (FBPM) \cite{Eis}, simulating optical pulses propagation in
nonlinear media by solving the NLSE
\begin{equation}\label{eqsim}
\frac{1}{2}\nabla ^2A + \left(\frac{2\pi}{\lambda}\right)^2\cdot
n\cdot \Delta n(x) A + i\left(\frac{2\pi n}{\lambda}\right)
\frac{\partial A}{\partial z} + \left( \frac{2
\pi}{\lambda}\right)^2\cdot n \cdot n_{NL}|A|^2A = 0
\end{equation}
was used. Here $A$ is the electric field amplitude in the paraxial
approximation.

In order to get a full correspondence between Eq. (\ref{eqsim}) and
the GPE (\ref{GPE}), we considered the time independent process in
the two dimensional $x,z$ space. The light beam propagates along the
$z$ axis so that the $z$ coordinate plays the role of time in the
GPE. $x$ is the 1-d traversal direction and corresponds to the space
coordinate in the GPE. The 'potential trap' is actually a waveguide
due to the variation of refractive index along the $x$ axis which
can be fabricated by correspondingly varying the dopant
concentration. The deviation $\Delta n$ of the refraction index from
its global value has the form

\begin{equation}\label{trap}
\Delta n(x) = \Delta n_{max} \left( 1 -
\frac{C}{\cosh^2(\frac{x}{100} )} \cdot
\left(\frac{x}{100}\right)^2\right)
\end{equation}
where we chose $C = 2.28$ for the calculations. An initial Gaussian
pulse centered at $x = 0$, $z = 0$ inside the trap is set to
propagate in such a medium with the Kerr nonlinearity. The relevant
medium (e.g., for AlGaAs) parameters are refractive index $n=3.33$
at $x \to \infty $, the Kerr nonlinearity $n_2=3\cdot 10^{-16}
\frac{m}{watt}$, the maximal deviation of refractive index $\Delta
n_{max} = 0.001$, the waveguide width $W = 2000 \mu m$, and its
length $L=35,000 \mu m $. The initial input is a Gaussian shaped
pulse of the width $w=300 \mu m$ and intensities in the range 700 to
2000 watt. The wavelength is $\lambda=3 \mu m$.

\section{Results}
\subsection{Emergence of the Blip}

The simulation results for zero, negative and positive
nonlinearities are presented in Fig. \ref{blip}. We clearly see how
a blip is created and propagates with a constant velocity
independently of the nature of the nonlinearity, i.e. the blip shows
up also in the case of zero or repulsive interaction in BEC, or zero
or defocusing Kerr nonlinearity in optics. It surely exists in the
negative nonlinearity case (see Fig. \ref{blip} - focusing case),
i.e. attractive interaction or focusing, in which case the blip
under proper conditions may transform into a bright
soliton.\cite{ours}.

Another universal feature of this phenomenon is the velocity of the
blip propagation. As can be seen in Fig. \ref{blip} it moves with a
constant velocity (along straight lines). Moreover, the velocity
remains the same for different nonlinearities. Our analysis in Ref.
\cite{ours} yields equation
\begin{equation}
v_{blip}= v_{shock}= \sqrt{\frac{2U_0}{m}}
\end{equation}
for the velocity of the dispersive shock wave propagation. Here
$U_0$ is the height of the potential barrier in the BEC problem. In
optics it is translated into the dimensionless quantity
\begin{equation}
v_{blip}=\sqrt{\frac{2\Delta n_{max}}{n}}.
\end{equation}
We see an excellent agreement between the theory and simulations
such as $v_{theory} = 1.4$ and $v_{simulation} = 1.3$ in the BEC
case and $v_{theory} = 2.45e^{-2} = 0.33$ and $v_{simulation} =
2.57e^{-2} = 0.35$ in optics regardless of the type of nonlinearity.
\begin{figure}\label{blip}
\begin{center}
\includegraphics[width=0.32\textwidth,height=0.25\textheight]{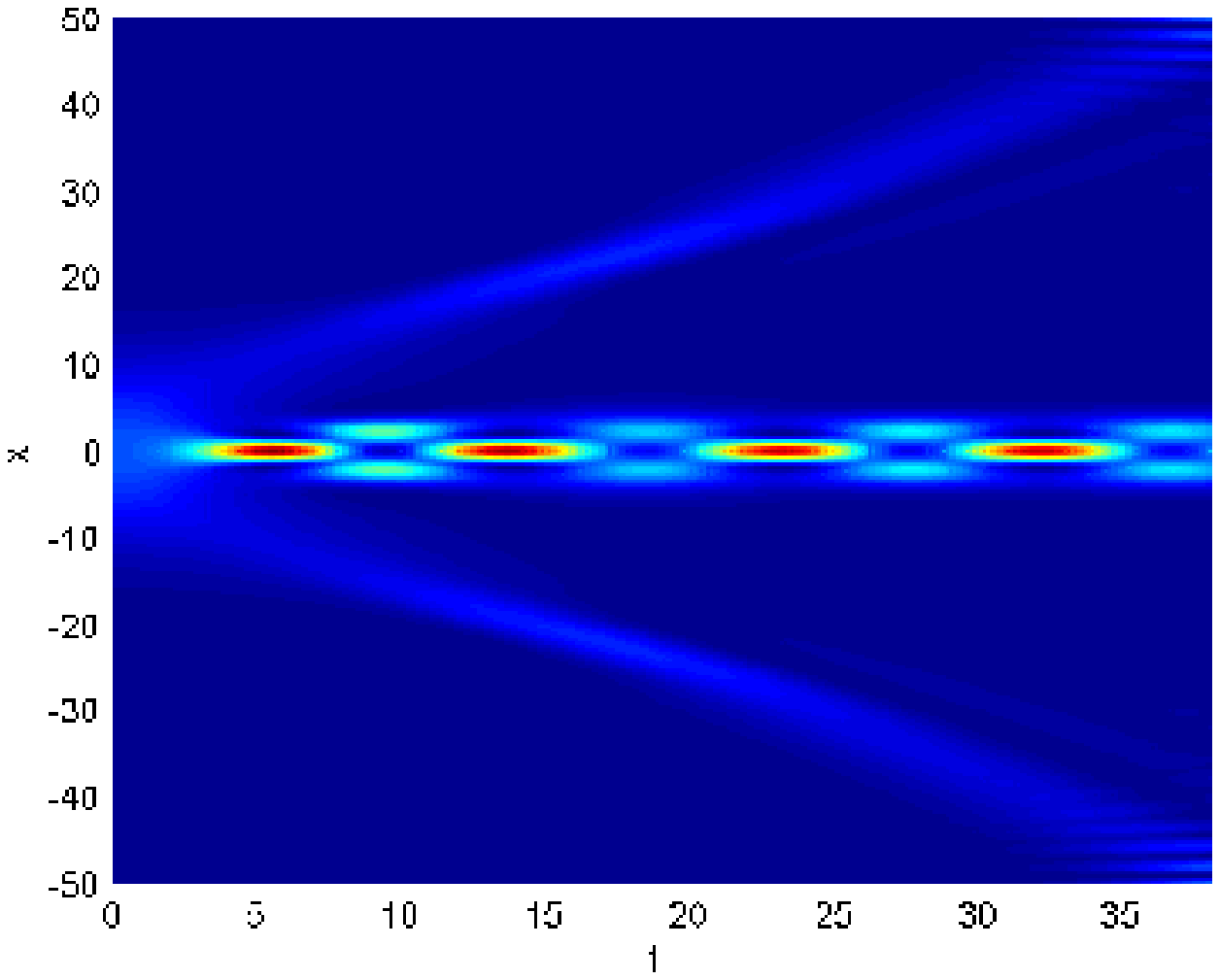}
\includegraphics[width=0.32\textwidth,height=0.25\textheight]{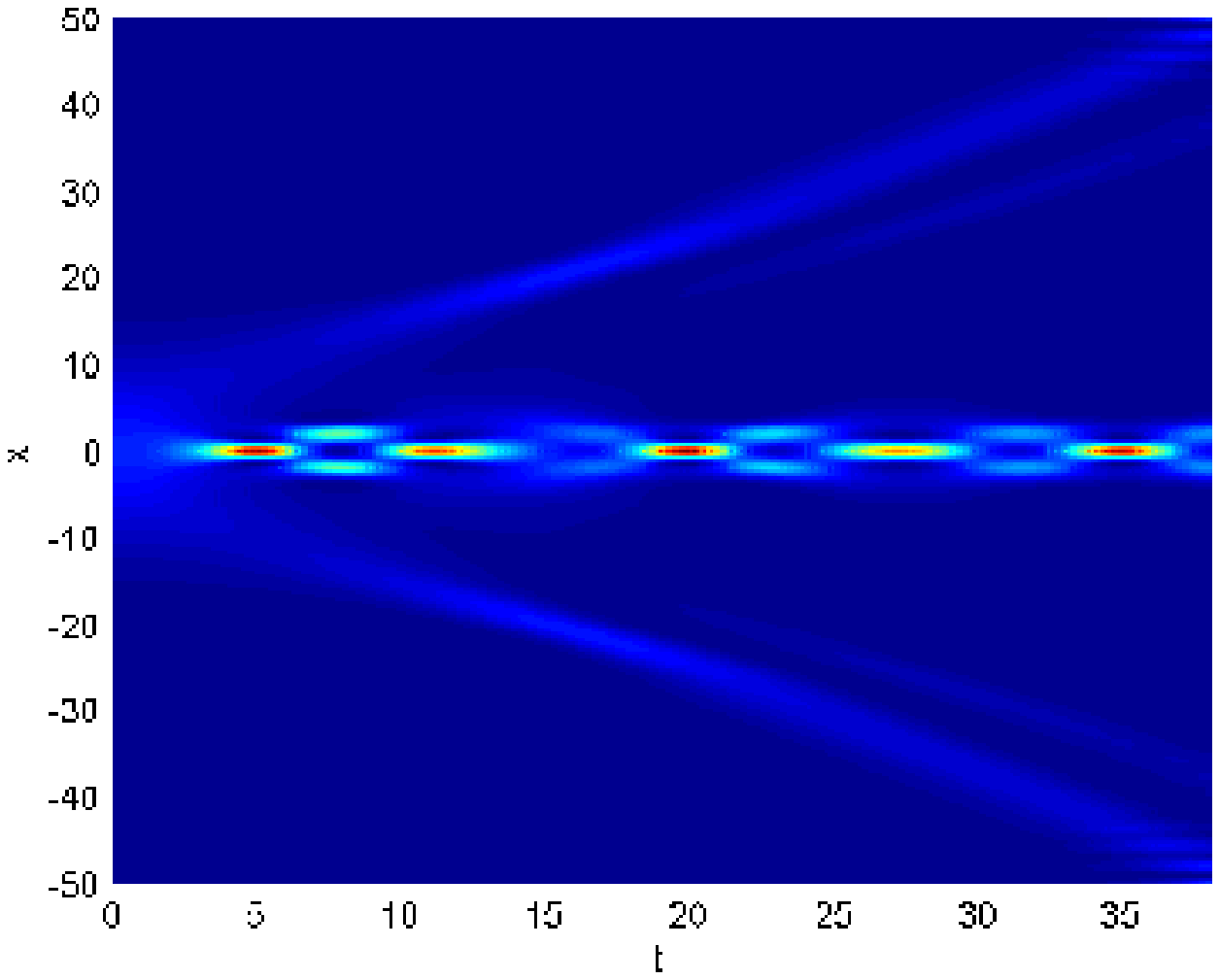}
\includegraphics[width=0.32\textwidth,height=0.25\textheight]{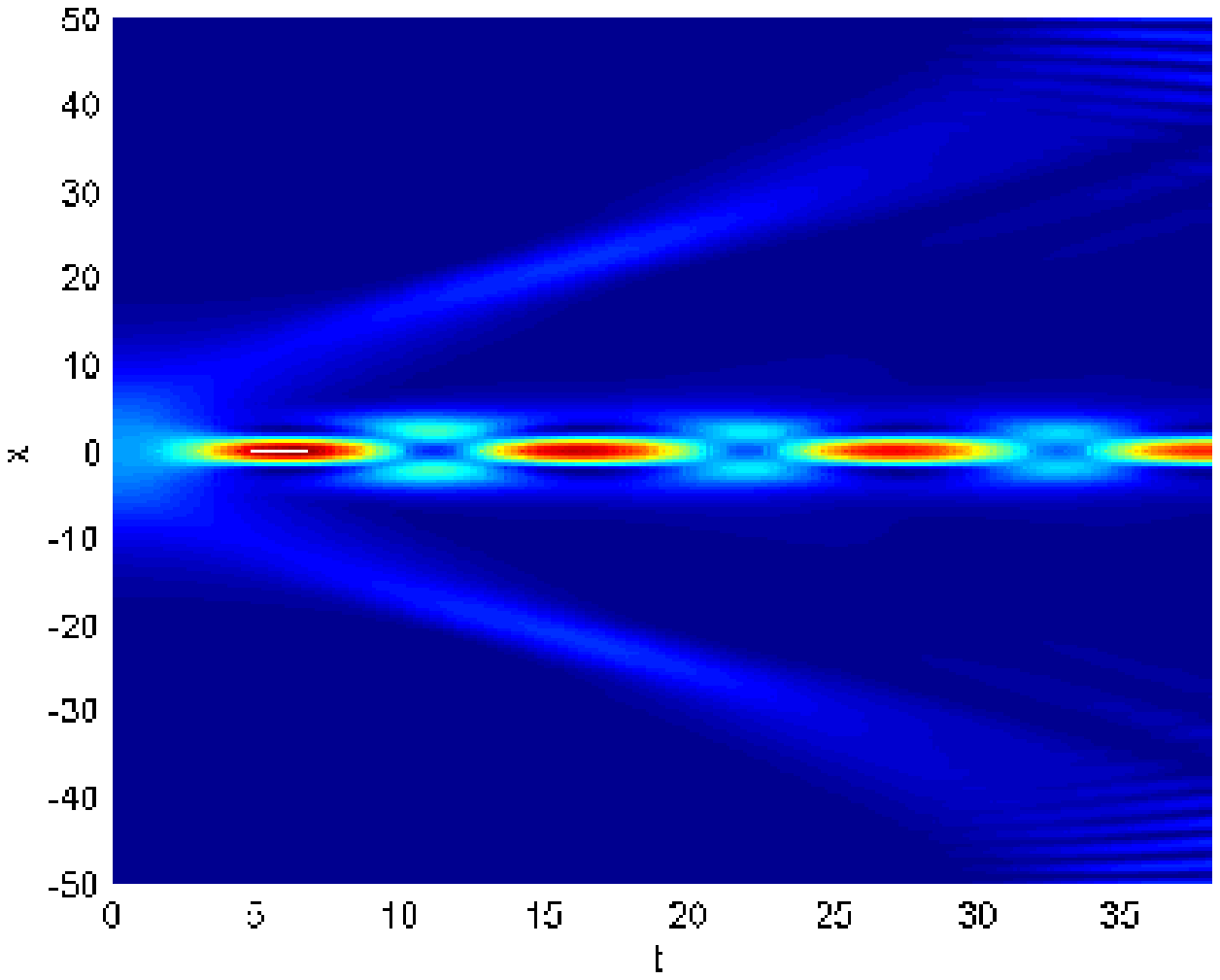}
\end{center}
\caption{Blip emergence and propagation. 
 The figures show continuous top view onto
the time evolution of BEC systems. From left to right: non focusing
(zero interactions), focusing (attractive interactions) and
de-focusing (repulsive interactions) cases . In the focusing case,
the blip transforms into a bright soliton. It is also clear 
 that the remaining trapped packet exhibit pulsations, a
feature that will be discussed below.}\label{blip}
\end{figure}

As was outlined in our previous paper\cite{ours} the blip results
from the tendency to a shock formation on the outskirts of the trap.
Fig. \ref{blip} shows how the focusing (attractive) nonlinearity
leads to a conversion of the pulse width, controlled by both
nonlinearity and dispersion, remains unchanged at least on the
timescale of our calculations. A special attention to formation of a
bright soliton from the emitted blip and its stability was given in
our previous paper \cite{ours}. In the two other cases, the
propagating blip starts exhibiting oscillatory tails typical of the
dispersive shock waves\cite{shock1,shock2,shock3,shock4}. These
oscillations are visibly stronger for the defocusing case. The shock
waves are governed by the dispersion and by defocusing media
(repulsive interaction in BEC).

\subsection{Pulsations inside the trap}

An interesting result is that the initial packet wider than the trap
very rapidly decays into a narrow packet loosely related to the
ground state inside the trap with high shoulders outside the trap.
This early time evolution takes place for all packets whose initial
width is wider than the trap.
%

\begin{figure}[tbp!]
\begin{center}
\includegraphics[width=8cm]{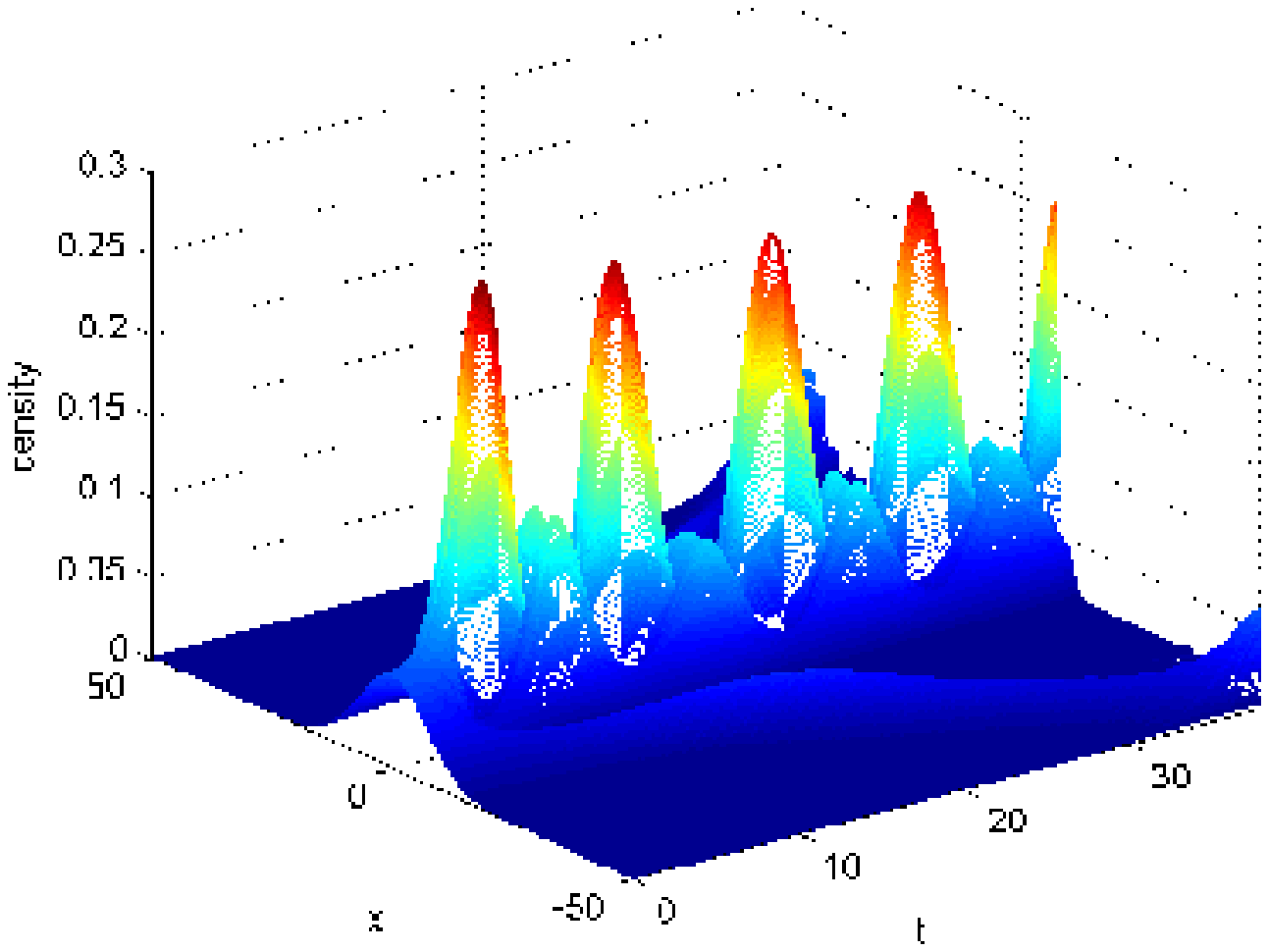}
\includegraphics[width=8cm]{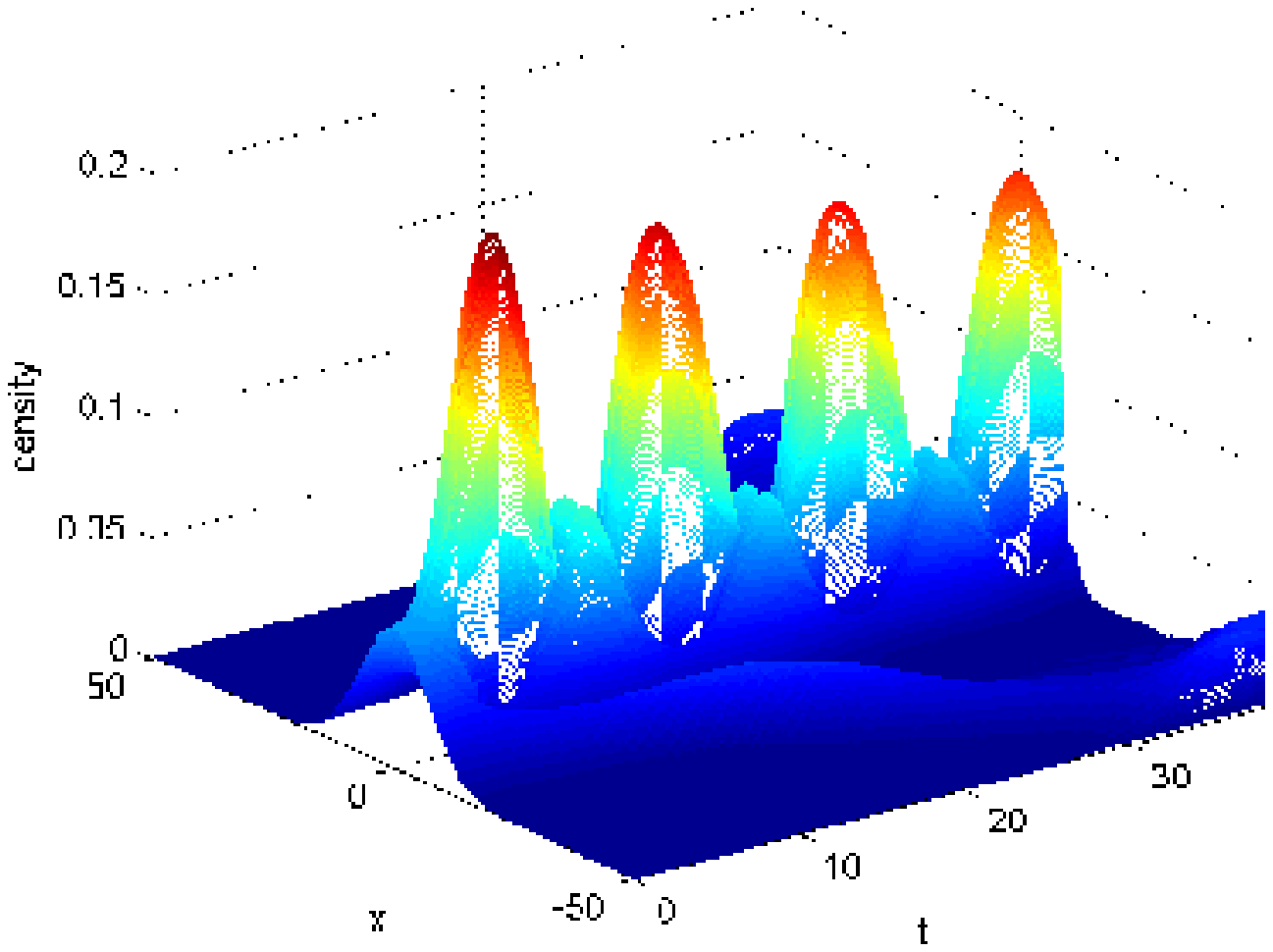}
\includegraphics[width=8cm]{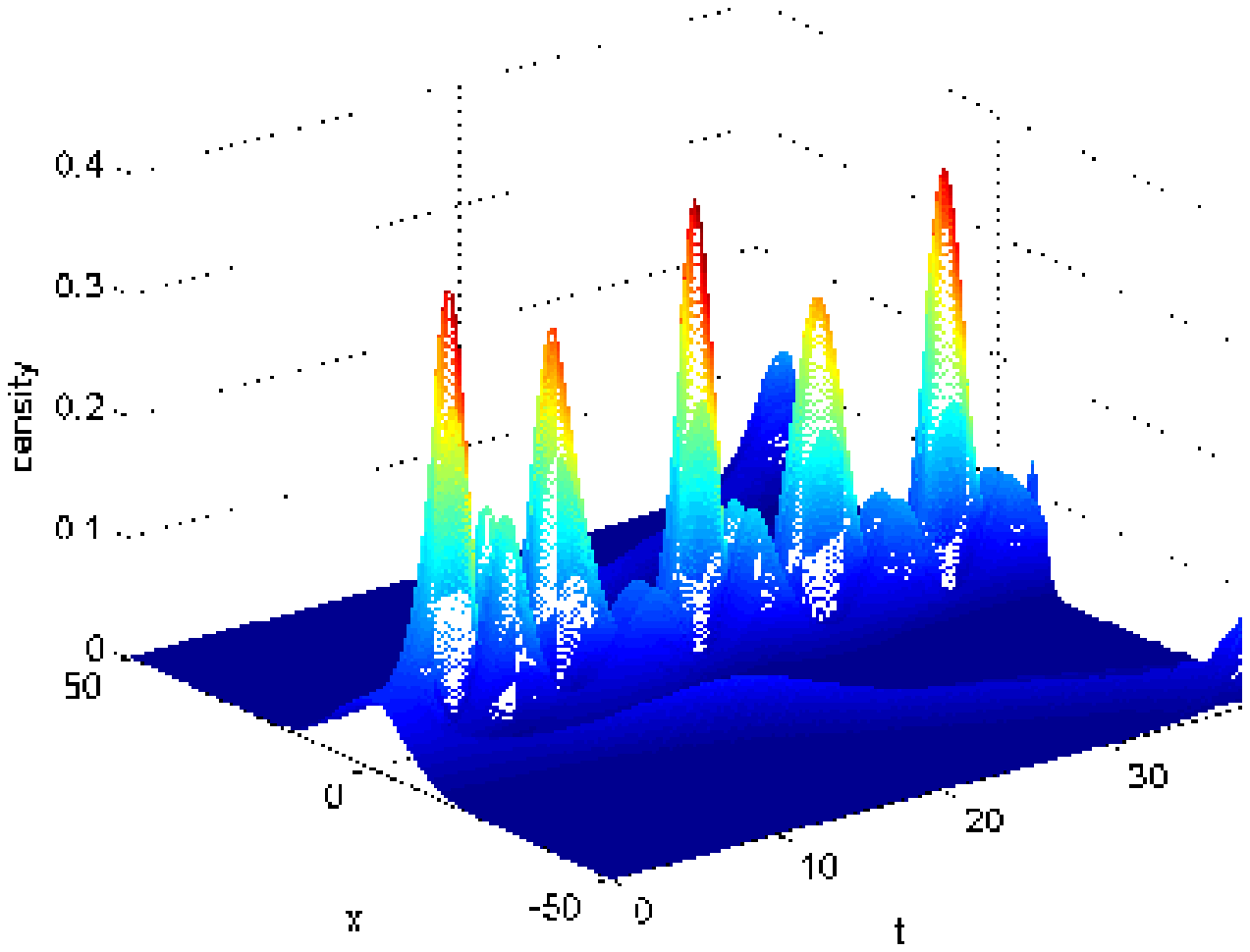}
\end{center}
\caption{Pulsations inside the trap after emission of a blip. From
top left: zero, positive (repulsive/defocusing) and negative
(attracting/focusing) nonlinearity.} \label{pulsations}
\end{figure}

As described above the initial packet emits a pair of blips and
forms a rather narrow packet inside the potential well, which is
rather close to the ground state of the well, obtained when
disregarding the tunneling. This narrow packet is a state with a
long life time which does not essentially decay during our
simulation time. However, we observe well pronounced pulsations of
the narrow packet with roughly doubled trap eigen frequency. These
pulsations are seen in Figs. \ref{blip} and are shown in the 3-d
plots in Fig. (\ref{pulsations}). We also see that in the focusing
case, the oscillations have higher frequency and the latter
increases as the nonlinearity increases, while in the defocusing
case (repulsive interaction in BEC) the pulsations are of lower
frequencies.

In order to analyze the pulsations within the trap we continue along
the lines of the hydrodynamic approach whose application to the
tunneling problem was presented in our previous paper\cite{ours}.
The NLSE/GPE for the complex function can be rewritten as two
equations for two real functions: the continuity equation
\begin{equation}\label{4}
\rho_t(x,t) + [\rho(x,t)v(x,t)]_x = 0.
\end{equation}
for the particle density distribution $\rho(x,t) = |\Psi(x,t)|^2$
and the Euler-type equation
\begin{equation}\label{5}
v_t(x,t) + v(x,t) v_x(x,t) = - \frac{1}{m} \nabla_x \left
[V_{ext}(x) - \frac{1}{\sqrt{\rho(x,t)}} \frac{\hbar^2}{2 m}
\nabla_x^2 \sqrt{\rho(x,t)} + \lambda \rho(x,t)\right].
\end{equation}
for the velocity field $v(x,t) = - \nabla_x \varphi(x,t)$ with
$\varphi(x,t)$ being the phase of $\Psi(x,t)$.

We assume the harmonic approximation
\begin{equation}\label{iniharm}
V_{ext}(x) = \frac{m\nu^2x^2}{2}
\end{equation}
for the shape of potential inside the trap and for the time being
neglect the nonlinearity. The density of narrow packet formed after
the emission of blips is assumed to have a Gaussian shape
$$
\rho_0(x) = \frac{N}{u_0\sqrt{\pi}} \exp\left \{- \frac{x^2}{u_0^2}
\right\}
$$
where $u_0$ does not necessarily coincide with the amplitude of zero
point oscillations $u_{ext} = \sqrt{\hbar/m\nu}$ in the potential
(\ref{iniharm}). It rather corresponds to another harmonic potential
with the frequency $\omega = \hbar/m u_0^2$.

We look for the solution of eqs. (\ref{4}) and (\ref{5}) also in the
Gaussian form
\begin{equation}\label{rho}
\rho(x) = \frac{1}{f(t)}\rho_0\left(\frac{x}{f(t)}\right)
\end{equation}
and
\begin{equation}\label{velo}
v(x,t) = x \frac{d}{dt}\ln f(t)
\end{equation}
where $u(t)=u_0 f(t)$ and $f(0)=1$. It means that the time
dependence of the solution is parameterized by the single function
$f(t)$. The functions (\ref{rho}) and (\ref{velo}) solve eq.
(\ref{4}) and when substituted into (\ref{5}) for $\lambda = 0$
yield equation
\begin{equation}\label{el}
f''= - \nu^2 f + \omega ^2 f^{-3}.
\end{equation}
for the function $f(t)$. This is the equation of motion for a
'particle' with the unit mass and with the coordinate $f$ in the
'effective potential'
\begin{equation}\label{effpot}
W_{eff}(f)= \nu^2 \frac{f^2}{2}+\frac{\omega^2}{2f^2}.
\end{equation}
The function $f(t)$ oscillates in this potential well between its
initial value $f_0 = f(0) = 1$ (we assume also that $f'(0) = 0$) and
the point $f_1 = \omega/\nu$ found from energy conservation
considerations. These two values of the function $f(t)$ correspond
to the minimal and maximal widths of the pulsating Gaussian packet.

To find the full solution for $f(t)$ we note that eq. (\ref{el}) can
be mapped onto the equation of motion for a radially symmetric 2d
harmonic oscillator with the frequency $\nu$ in which $f$ is its
radial coordinate. Then eq. (\ref{el}) is obtained for the motion
with the conserved angular momentum $\omega$.

The solution satisfying the above initial conditions is
\begin{equation}\label{f}
f(t) = \sqrt{\cos^2\nu t + \frac{\omega^2}{\nu^2} \sin^2 \nu t}.
\end{equation}
which describes the dynamics of the width of the trapped packet.

It is interesting to indicate that there is a more general
analytical solution
\begin{equation}\label{rho_2}
\rho(x) = \frac{1}{f(t)}b^2\left(\frac{x}{f(t)}\right)
\rho_0\left(\frac{x}{f(t)}\right)
\end{equation}
where
\begin{equation}\label{bs}
b^2\left(\frac{x}{f(t)}\right) = C_1 + i C_2 f(t) \mbox{erf}\left(
\frac{ix}{u_0f(t)}\right)
\end{equation}
with the same parameterizing function $f(t)$ as in (\ref{rho}). The
parameters $C_1$ and $C_2$ should be found from the normalization of
the density distribution (\ref{rho_2}) and its initial shape. The
error function in (\ref{bs}) is odd in $x$ which makes the
distribution (\ref{rho_2}) asymmetric unless we assume that one of
the parameters $C_1$ or $C_2$ zeros. If $C_2 = 0$ we return to the
Gaussian distribution considered above. If, however, $C_2 =0$ we
have a symmetric distribution with a minimum at $x=0$ similar to
what we see at the maximum amplitude of the pulsations.

In the presence of nonlinearity ($\lambda \neq 0$) we cannot get an
exact solution, however we may consider eqs. (\ref{rho}) and
(\ref{velo}) as an approximate trial functions and find the energy
of the pulsating packet in units $\hbar\nu$,
\begin{equation}\label{dropletE}
\varepsilon_{drop} = \frac{1}{2} \dot f^2 + W_{eff}(f)
\end{equation}
where the first term in the r.h.s. stands for the kinetic energy of
the pulsations. The second term
\begin{equation}\label{effpot-n}
W_{eff}(f)= \nu^2\frac{f^2}{2} + \omega^2 \frac{1}{2f^2} +
\frac{\lambda N}{f\sqrt{2\pi}}
\end{equation}
represents the new effective potential (instead of (\ref{effpot})),
which is just the internal energy of the packet at the given radius
$fu_0$ obtained by averaging the BEC hamiltonian (\ref{GPE}) with
the above trial function. It means that the oscillations of the
packet are governed by the equation of motion
$$
f''= - \nu^2 f + \omega ^2 f^{-3} + \frac{\lambda
N}{f^2\sqrt{2\pi}}.
$$

\begin{figure}
\includegraphics[width=8cm,angle=-90]{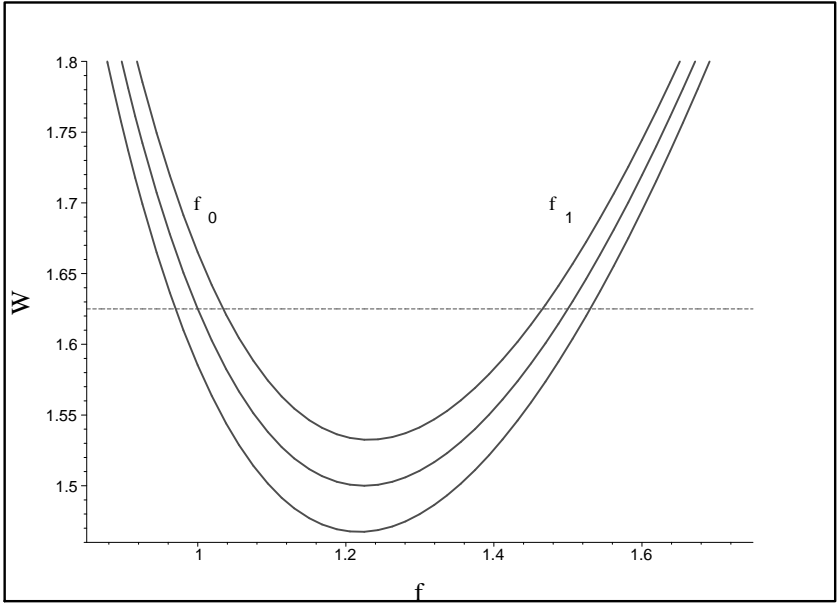}
\caption{The effective potential (\ref{effpot-n}) for various
interaction parameters: $N \lambda/\nu^2= 0.1$ --- upper curve, $N
\lambda/\nu^2 = 0$ --- middle curve, $N \lambda/\nu^2 = -0.1$ ---
lower curve. $\omega/\nu = 1.5$. The horizontal line shows the level
of oscillations between the points $f_0$ and $f_1$ in the case of
$\lambda = 0$. }\label{pot3}
\end{figure}

Fig. \ref{pot3} presents the effective potentials for three values
of the dimensionless nonlinearity $N\lambda/\nu^2$. We see in this
figure that the repulsive interaction decreases the minimal width
and increases the maximal width, i.e. it increases the amplitude of
pulsations. A more detailed analysis shows that the period of
pulsations also increases --- the frequency lowers. The attractive
interaction acts in the opposite direction and the frequency grows.
This type of behavior agrees with the simulation results which can
be clearly seen from Fig. \ref{pulsations} - the number of peaks in
the same time interval varies for the three graphs corresponding to
the different values of the interaction parameter.

The graph in fig. \ref{pulsations} for the nonfocusing case (zero
interaction) shows that the period is roughly $T = 8$. To estimate
the trap frequency, which is nearly harmonic near the center of the
well, we fit it to a parabolic function from which we deduce the
frequency to be around $\nu_{trap}\simeq 0.0737$. Therefore the
expected period would be $T = \pi/\nu_{trap} \simeq 8.18$\ in a very
good agreement with the theoretical prediction.

The simulation results for the NLSE, eq. (\ref{eqsim}), show that
the fraction of the initial pulse which stays trapped along the
propagating axis, exhibits pulsations of period $ \sim 5000 m\mu$.
To discuss the frequency equivalent for light propagation we look at
the second term in eq. (\ref{eqsim}). Taking only the inside part of
the trap and removing the constant offset, this term can be
rewritten as
\begin{equation}
\sim -\left(\frac{2\pi}{\lambda}\right)^2\cdot n\cdot  \Delta
n_{max} \cdot \left(\frac{x}{100}\right)^2
\end{equation}
and is to be fitted to the harmonic potential $\nu^2 x^2/2$ for
trapped particles. Then we get $ T = \pi/\nu = 7200 \mu$m for 'the
spatial period', which is close to the simulation result
$T_{sim}\sim 5000 \mu$m.

It is worth mentioning here that various type of oscillations have
been discussed in literature (see, review \cite{stringani}) as a
rule using the Thomas-Fermi approximation which neglects the quantum
pressure term. This approximation works especially well for small
amplitude oscillations with the repulsive interaction when the
density distribution is rather close to uniform. It is emphasized
here that the density distribution in the tunneling problem we
discuss here is strongly nonuniform and the pulsations discussed
here are strong so that the role of the quantum pressure is of an
utter importance.

\section{Atomic Pulsed Laser}

Combining the two discussed effects, i.e. blip emission due to
macroscopic tunneling and in-trap periodic pulsations, one can
propose a new mechanism of controllable multiple blip emission. Such
effect may lead to a realization of Atomic Pulsed Laser. The
mechanism we propose is based on narrowing the well to a desired
width, according to eq. (\ref{effpot}) at times that match the
pulsating packet being at its maximum width and allows for
outcoupling matter waves pulses. In addition to the simple
realization, a full control and modulation possibility of several
solitonic features, also in the course of emission, is at hand. In
order to demonstrate this we consider the potential
\begin{equation}
U_{ext} = C(t) \ \frac{x^2}{\cosh^2( \frac{x}{a(t)})}.
\end{equation}
with the varying in time parameters
\begin{equation}
a(t),C(t)=\left\{\begin{array}{l} a_1,C_1\quad t<t_1
\\a_2,C_2\quad t<t_2\\...\end{array}\right.
\end{equation}
Variations of $a(t)$ and $C(t)$ are fitted in such a way as to keep
the height of the barrier unaltered if identical velocities of the
pulses are desired. Alternatively they may be chosen to vary the
barrier height in order to reach the desired variation of the
velocities of outcoupled pulses. $a(t)$ is the variable width of the
well. The times $t_1$, $t_2$ ... are set to match the period of
pulsations. It is important to stress here that the latter changes
according to the width of the narrowing trap as was proved earlier.
Therefore, on-switch times alter accordingly.

\begin{figure}
\includegraphics[width=12cm,angle=0]{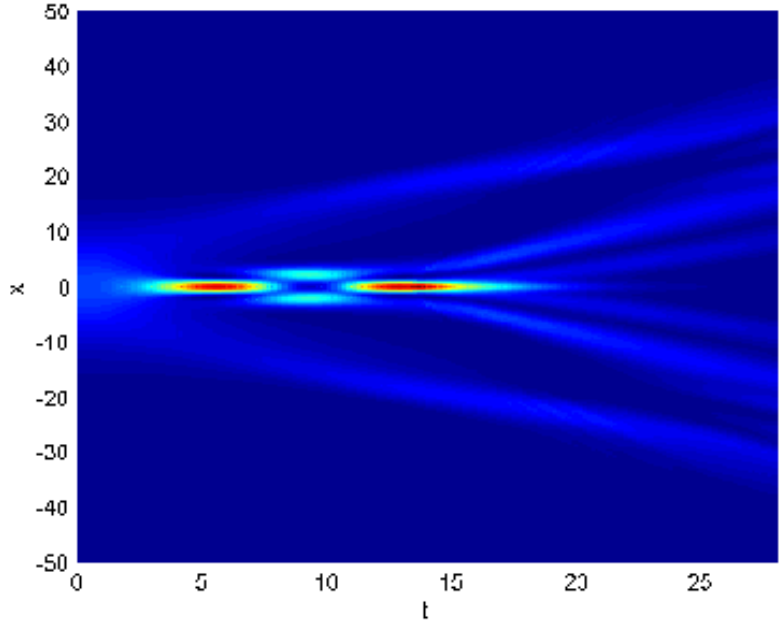}
\caption{Emission of the second and third pairs of blips by the trap
twice subsequently squeezed at the maximal width of
pulsations.}\label{pot_time}
\end{figure}

Three pairs of blips emitted from the trap are seen in fig.
\ref{pot_time}. The first pair of blips is emitted at the start of
the process and the two other pairs are emitted due the two
narrowings at $t = 8$ and $t = 13$ with the widths $a=5$, $4$ and
$0.7$ respectively. Since the nonlinearity plays a minor role in the
emission process this simulation is carried out without it.

\begin{figure}
\begin{center}
\includegraphics[width=6.5cm]{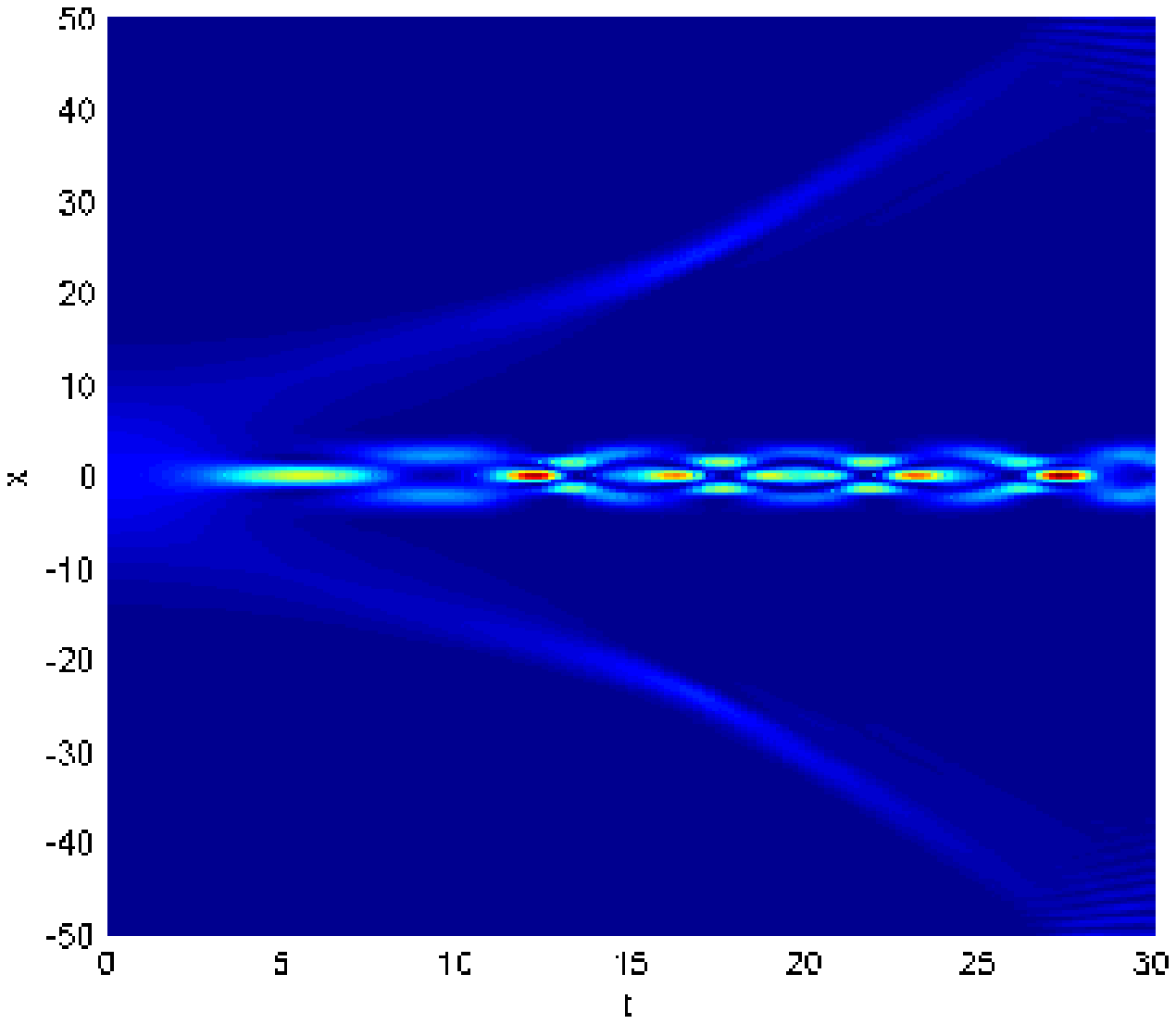}
\includegraphics[width=7cm]{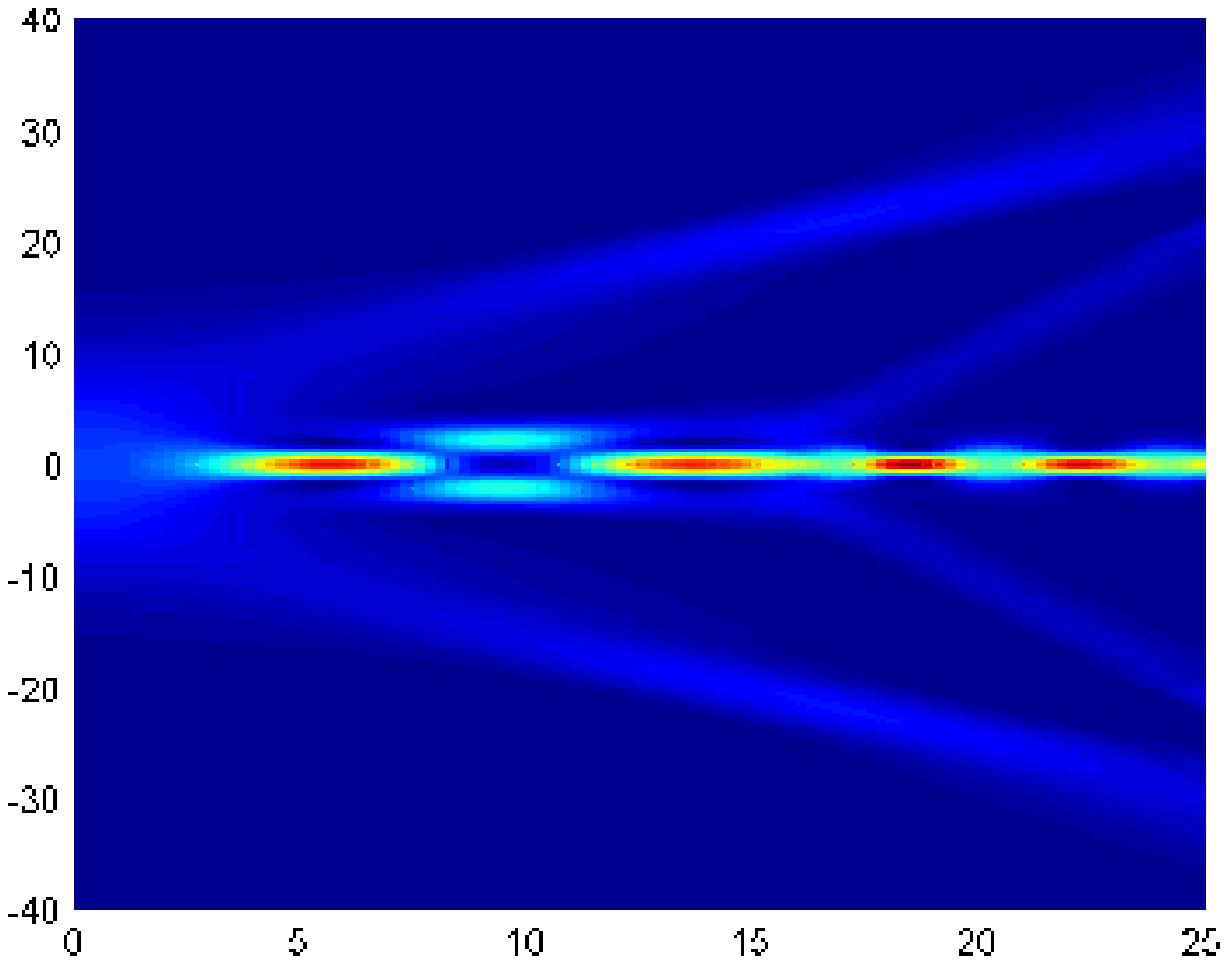}
\end{center}
\caption{Velocity modulation achieved by raising the barrier height.
Left: modulation of single pulse velocity, during its propagation.
At $t=15$ the barrier peak is raised from $\simeq 1$ to $\simeq 4$
Right: production of two consecutive pairs of pulses with different
velocities. At $t=15$ the trap is narrowed from a=5 to a=2.5 and the
barrier peak is raised from $\simeq 1$ to $\simeq 2$ }\label{vel}
\end{figure}

{\bf Velocity modulation}. As stated above, our analysis, based on
the hydrodynamic approach, predicts a one to one relation between
the velocity and the height of the barriers, leading to a viable
control of the outcoupled pulse velocities also in the course of
sequential emissions. In this case the velocity of the blips is
altered according to
\begin{equation}
v_{blips}=\sqrt{2C(t)s(t)}
\end{equation}
where $s(t)$ is the solution of the transcendental equation
\begin{equation}
s=\frac{\tanh\left(a(t)\ s\right)}{a(t)}
\end{equation}
Fig. \ref{vel} shows the velocity variation for a single pair of
pulses during its propagation, and for two pairs of pulses which are
made to propagate with different velocities.

The velocity difference between the two consecutive outcoupled
pulses has been created by temporally changing the barrier height.
The restriction here being that the first propagating outcoupled
pulse has to be far enough from the barrier region, so that it will
not be effected as well.

{\bf Temporal interval modulation and control} Temporal spacing
modulation can be achieved by considering different numbers of
pulsation periods between two emissions. Another feature that should
be taken into consideration is the time of the blip formation. This
time is shown in ref.\cite{ours} to vary according to the trap
parameters as
\begin{equation}
t_b\sim\left(\frac{8 U_0}{a^2} \right)^{-\frac{1}{2}}
\end{equation}
Therefore the formation time becomes shorter for the narrower
barriers.

{\bf Stability} The pulse creation is not related to the
nonlinearity of the GPE and the present discussion without loss of
generality is limited to the zero interaction. However, it is of
course evident that for an operating pulsed laser the pulses have to
transform into bright solitons and moreover need to be kept stable
and not disperse. This can be achieved by turning on attractive
interaction along the way with lower limit according to the soliton
formation condition
\begin{equation}\label{stab}
U_{int}\cdot|\psi(0,T)|^2\sim 6.5 a^{-2} + 0.03 U_0
\end{equation}
found in ref. \cite{ours} for the shape of trap potential which is
rather close to the trap chosen here. In fact the parameter values
in eq. (\ref{stab}) may slightly change with the shape of the
particular trap. Here $T$ is the on-set time. $|\psi(0,T)|^2$ the
density in the center of the trap at $t=T$. If the condition
\ref{stab} is satisfied, the emerging pulse may transform into a
stable bright soliton. In 2d and 3d case some stricter conditions as
say Vakhitov-Kolokolov\cite{VK73} should be satisfied. Since the
emerging pulses are much more dilute than the initial packet,
stability against collapse in 2d and 3d case is easier achievable.
The masses and widths of the pulses can be also found as described
for the single emission event in ref. \cite{ours}.

In principle we can consider restoring the original width of the
trap after each emission and repeat this process as many times as
necessary. This mechanism will work as long as there is matter left
within the trap and its coherency is held.

\section{Conclusions}

Certain aspects of the universality of the emission of blips in the
course of macroscopic tunneling were asserted, using two simulation
programs, one of which applying directly to Optics. Pulsations of
the trapped packet after emergence of the blips were shown by the
hydrodynamic approach and by numerics to occur with the period
matching twice the trap frequency, which inspired us to suggest a
mechanism for realization of Atom Pulsed Laser. The blips of the
matter density may be emitted by controllably narrowing the trap.
The velocity of the outgoing pulses may be controlled by varying the
height of the barrier. We show also how to control the times of the
pulse emissions and their masses.

{{\bf Acknowledgments.}\ \  The authors acknowledge the support of
United States - Israel Binational Science Foundation, Grant N
2006242 and of Israeli Science Foundation, Grant N 944/05. The
authors are also grateful to Max Planck Institute for Physics of
Complex Systems, Dresden, for hospitality. We are grateful to S.
Bar-Ad, Y. Linson, and M. Weitz for help and stimulating
discussions.}


\begin{thebibliography}{99}


\bibitem{becdark1} S. Burger, K. Bongs, S. Dettmer, W. Ertmer,
K. Sengstock, Phys. Rev. Lett., {\bf 83}, 5198 (1999).

\bibitem{becdark2} J. Denschlag, J.E. Simsarian, D.L. Feder,
C.W. Clark, L.A. Collins, J. Cubizolles, L. Deng, E.W. Hagley, K.
Helmerson, W.P. Reinhardt, S.L. Rolston, B.I. Schneider, W.D.
Phillips, Science, {\bf 287}, 97 (2000).

\bibitem{becsol1} L. Khaykovich, F. Schreck, G. Ferrari, T.
Bourdel, J. Cubizolles, L.D. Carr, Castin, Y., C. Salomon, Science
{\bf 296}, 1290 (2002).

\bibitem{becsol2} K.E. Strecker, G.B. Partridge, A.G. Truscott,
R.G. Hulet, Nature (London) {\bf 417}, 150 (2002).

\bibitem{atop1}  M.O. Mewes, M.R. Andrews, D.M. Kurn, D.S. Durfee,
C.G. Townsend, W. Ketterle, Phys. Rev. Lett., {\bf 78}, 582 (1997).

\bibitem{atop2} B.P. Anderson, and M.A. Kasevich, Science,
{\bf 282}, 1686 (1998).

\bibitem{atop3} I. Bloch, T.W. Hansch, and T. Esslinger,
Phys. Rev. Lett., {\bf 82}, 3008 (1999).

\bibitem{atop4} E.W. Hagley, L. Deng, M. Kozuma, J. Wen, K.
Helmerson, S.L. Rolston, W.D. Phillips, Science, {\bf 283}, 1706
(1999).

\bibitem{atop5} J.L. Martin, C.R. McKenzie, N.R. Thomas, J.C. Sharpe,
D.M. Warrington, P.J. Manson, W.J. Sandle, A.C. Wilson, J. Phys. B,
{\bf 32}, 3065 (1999).

\bibitem{pulsedlaser1} M.I. Rodas-Verde, H. Michinel, V.M. P\'{e}rez
Garc\'{\i}a, Phys. Rev. Lett., {\bf 95}, 153903 (2005).

\bibitem{pulsedlaser2} A.V. Carpentier, H. Michinel, M.I.
Rodas-Verde, V.M. P\'{e}rez Garc\'{\i}a, Phys. Rev. A, {\bf 74},
013619 (2006).

\bibitem{pulsedlaser3} L.D. Carr, J. Brand, Phys. Rev. A, {\bf 70},
033607 (2004).

\bibitem{pulsedlaser4} P.Y.Y Chen, B.A. Malomed, J. Phys. B: At. Mol. Opt. Phys.
{\bf 38} 4221 (2005).

\bibitem{pulsedlaser5} P.Y.Y Chen, B.A. Malomed, J. Phys. B.: At. Mol. Opt.
Phys. {\bf 39} 2803 (2006).

\bibitem{fetter} A.L. Fetter, Phys. Rev. Lett., {\bf 138}, A429
(1965).

\bibitem{stringani} F. Dalfovo, S. Giorgini, L.P. Pitaevskii, S.
Stringari, Rev. Mod. Phys. {\bf 71}, 463 (1999).

\bibitem{anal} E.A. Ostrovskaya, Y.S. Kivshar, M. Lisak, B.
Hall, F. Cattani, D. Anderson, Phys. Rev. Lett. {\bf 61}, 031601
(2000).

\bibitem{chm05} L.D. Carr, M.J. Holland, B.A. Malomed, J.Phys B
{\bf 38}, 3217 (2005).

\bibitem{mcmb04} N.Moiseyev, L.D. Carr, B.A. Malomed, and Y.B. Band, J.
Phys. B {\bf 37}, L193 (2004).

\bibitem{om05} S. Osovski and N. Moiseyev, Phys. Rev. A {\bf 72}, 033603 (2005)

\bibitem{made} E. Madelung, Z. Phys. {\bf 40}, 322 (1927).

\bibitem{marbu} J.H. Marburger, Prog. Quantum Electron. {\bf 4}, 35 (1975).

\bibitem{silber} Y. Silberberg, Opt. Lett. {\bf 15}, 1282 (1990).

\bibitem{ablo} M.A. Hoefer, M.J. Ablowitz, I. Coddington, E.A. Cornell, P.
Engels, and V. Schweikhard, Phys. Rev. A {\bf 74}, 023623 (2006).

\bibitem{Holland} P. R. Holland, 1993 {\em The Quantum Theory of Motion} Cambridge
Univ. Press

\bibitem{stringani2} S. Stringani, Phys. Rev. Lett. {\bf 77}, 2360
(1996).

\bibitem{LF01} M. Levanda and V. Fleurov, Annals of Physics, {\bf 292}, 199 (2001)

\bibitem{fs05} V. Fleurov and A. Soffer, Europhys.Lett., {\bf 72},
287 (2005).

\bibitem{ours} G. Dekel, V. Fleurov, A. Soffer, C. Stucchio,
Phys. Rev. A., {\bf 75}, 043617 (2007).

\bibitem{tunntime1}  Ph. Balcou, L. Dutriaux, Phys. Rev.
Lett., {\bf 78}, 851 (1997).

\bibitem{tunntime2} E. Pollak, W.H. Miller, Phys, Rev.
Lett., {\bf 53}, 116 (1984).

\bibitem{tunntime3} M. B\"uttiker, R. Landauer, Phys. Rev.
Lett., {\bf 49}, 1740 (1982).

\bibitem{tunntime4} E.H. Hauge, J.A. St{\o}vneng, Rev. Mod.
Phys., {\bf 61}, 917 (1989).

\bibitem{shock1} M.A. Hoeffer, M.J. Ablowitz, I. Coddington,
E.A. Cornell, P. Engels, V. Schweikhard, Phys. Rev. A., {\bf 74},
023623 (2006).

\bibitem{shock2} B. Damski, Phys. Rev. A., {\bf 69}, 043610 (2004).

\bibitem{shock3} A.M. Kamchatnov, A. Gammal, R.A. Kraenkel, Phys.
Rev. A., {\bf 69}, 063605 (2004).

\bibitem{shock4} G.A. El, A.M. Kamchatnov, Phys. Lett. A.,
{\bf 350}, 192 (2006).

\bibitem{shock5} Wenjie Wan, Shu Jia and J. W. Fleischer, Nature
Physics {\bf 3}, 46 - 51 (2007).

\bibitem{Ballagh} R.J. Ballagh, Computation methods for solving
nonlinear partial differential equations. University of Otago
(2000).

\bibitem{Eis} The program was prepared by H. Eisenberg, Hebrew
University of Jerusalem.

\bibitem{VK73}  N. G. Vakhitov and A. A. Kolokolov, Izv. Vyssh. Uch.
Zaved., Radiofizika {\bf 16}, 10 120 (1973) [Radiophys. Quantum
Electron. {\bf 16}, 783 (1973)]

\end{thebibliography}
\end{document}